\documentclass{article}
\def\Ref#1{(\ref{#1})}
\usepackage{amsmath}
\usepackage{cite}
\begin{document}
\begin{titlepage}
\noindent{\large \bf Exactly solvable models through the empty
interval method}

\vskip 2 cm

\begin{center}{M. Alimohammadi$^a${\footnote
{alimohmd@theory.ipm.ac.ir}} , M. Khorrami$^b${\footnote
{mamwad@iasbs.ac.ir}}\& A. Aghamohammadi$^c${\footnote
{mohamadi@theory.ipm.ac.ir}} } \vskip 5 mm

{\it{ $^a$ Physics Department, University of Tehran,
    North Karegar Ave.,\\ Tehran, Iran}

{ $^b$ Institute for Advanced Studies in Basic Sciences,
             P.~O.~Box 159,\\ Zanjan 45195, Iran. }

{ $^c$ Department of Physics, Alzahra University,
             Tehran 19834, Iran. }}
\end{center}

\begin{abstract}
\noindent The most general one dimensional reaction-diffusion
model with nearest-neighbor interactions, which is
exactly-solvable through the empty interval method, has been
introduced. Assuming translationally-invariant initial conditions,
the probability that $n$ consecutive sites are empty ($E_n$), has
been exactly obtained. In the thermodynamic limit, the large-time
behavior of the system has also been investigated. Releasing the
translational invariance of the initial conditions, the evolution
equation for the probability that $n$ consecutive sites, starting
from the site $k$, are empty ($E_{k,n}$) is obtained. In the
thermodynamic limit, the large time behavior of the system is also
considered. Finally, the continuum limit of the model is
considered, and the empty-interval probability function is
obtained.

\end{abstract}
\begin{center} {\bf PACS numbers:} 82.20.Mj, 02.50.Ga, 05.40.+j

\end{center}

\end{titlepage}
\newpage
\section{Introduction}
The principles of equilibrium statistical mechanics are well
established. But thermal equilibrium is a special case, and little
is known about the properties of systems not in equilibrium, for
example about the relaxation toward the stationary state. There is
no general approach to systems far from equilibrium. As mean-field
techniques, generally, do not give correct results for
low-dimensional systems, people are motivated to study
exactly-solvable stochastic models in low dimensions. Moreover,
solving one-dimensional systems should in principle be easier.
Different methods have been used to study these models, including
analytical and asymptotic methods, mean-field methods, and
large-scale numerical methods. Exact results for some models on a
one-dimensional lattice have been obtained, for example in
\cite{ScR,ADHR,KPWH,HS1,PCG,HOS1,HOS2,AL,AKK,RK,RK2,AKK2,AAMS,AM1,MA,RK3}.

The term exactly-solvable have been used with different meanings.
In \cite{GS}, a ten-parameter family of reaction-diffusion
processes was introduced for which the evolution equation of
$n$-point functions contains only $n$- or less- point functions.
The average particle-number in each site has been obtained exactly
for these models. In \cite{AM2}, the same method has been used to
analyze the above mentioned ten-parameter family model on a finite
lattice with boundaries. In \cite{AA} and \cite{RK3},
integrebility means that the $N$-particle conditional
probabilities' S-matrix is factorized into a product of 2-particle
S-matrices.

The empty interval method (EIM) has been used to analyze the one
dimensional dynamics of diffusion-limited coalescence
\cite{BDb,BDb1,BDb2,BDb3}. Using this method, the probability that
$n$ consecutive sites are empty has been calculated. This method
has been used to study a reaction-diffusion process with
three-site interactions \cite{HH}. EIM has been also generalized
to study the kinetics of the $q$-state one-dimensional Potts model
in the zero-temperature limit \cite{Mb}.

In this article, we are going to study all the one dimensional
reaction-diffusion models with nearest neighbor interactions which
can be exactly solved by EIM. It is worth noting that ben-Avraham
et al. have been studied one-dimensional diffusion-limited
processes through EIM \cite{BDb,BDb1,BDb2,BDb3}. In their study,
some of the reaction rates have been taken infinite, and they have
worked out the models on continuum. For the cases of finite
reaction-rates, some approximate solutions have been obtained.

We study models with finite reaction rates, obtain conditions for
the system to be solvable via EIM, and then solve the equations of
EIM. We do this for a system on a lattice and on continuum.

The scheme of the paper is as follows. In section 2, the most
general one dimensional reaction-diffusion model with
nearest-neighbor interactions which can be solved exactly through
EIM has been introduced. Assuming translational invariance, the
probability that $n$ consecutive sites are empty ($E_n$), has been
exactly obtained. In the thermodynamic limit, the large-time
behavior of the system has also been investigated. In section 3,
the assumption of translational invariance has been released, and
the evolution equation for the probability that $n$ consecutive
sites, starting from the site $k$, are empty ($E_{k,n}$) is
obtained. In the thermodynamic limit, the large time behavior of
the system is also considered. It is shown that
translationally-asymmetric fluctuations relative to the stationary
configuration disappear faster than the translationally-symmetric
fluctuations. In section 4, the continuum limit of the model is
considered, and the empty-interval probability function is
obtained.

\section{Models solvable through the empty interval method:
the translationally-invariant case} Consider a general one-species
reaction-diffusion model on a one-dimensional lattice with $L+1$
sites, with nearest-neighbor interactions. We want to impose
restrictions on the reaction- and diffusion-rates, so that the
system is solvable via EIM, that is, so that the evolution
equation for the probability that $n$ consecutive sites are empty
($E_n$) is closed. Suppose that the initial condition of the
system is translationally-invariant. The most general interactions
for a single species model in a one-dimensional lattice with
nearest-neighbor interactions are
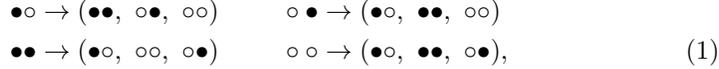
\begin{align}\label{0}
  \bullet\circ&\to (\bullet \bullet,\  \circ\bullet,\   \circ\circ
  )\qquad\
  \circ\bullet\to(\bullet\circ,\    \bullet \bullet,\  \circ\circ
  )\nonumber \\
  \bullet \bullet&\to (\bullet\circ,\  \circ\circ,\ \circ\bullet
  )\qquad\
  \circ\circ\to(\bullet\circ,\   \bullet \bullet,\  \circ\bullet ),
\end{align}
where an empty (occupied) site is denoted by $\circ$ ($\bullet$).
The constraint of solvability of the model through EIM, imposes
that, as we will show, there are no processes in which the final
configuration is $\circ\circ$. We shall also see that the
processes the initial configuration of them is $\circ\circ$, have
no effect on the solvability through EIM. But first, let us
consider only the systems for them there are no interactions with
$\circ\circ$ as the initial or final configuration. So, among the
above 12 interactions, only the following 6 interactions remain to
be considered.
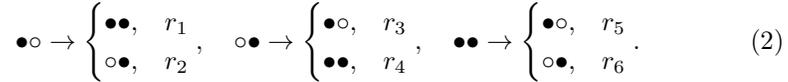
\begin{equation}\label{e1}
  \bullet \circ \to \begin{cases}\bullet \bullet,& r_1\\
                   \circ \bullet,& r_2
                   \end{cases},
\quad  \circ \bullet \to\begin{cases} \bullet \circ,& r_3\\
                               \bullet \bullet,& r_4
                 \end{cases},
\quad  \bullet\bullet   \to \begin{cases}\bullet\circ,& r_5\\
                        \circ \bullet,& r_6
            \end{cases}.
\end{equation}
The parameters $r_i$ are the rate of interactions. Define
\begin{equation}\label{2}
P(\overbrace{\circ\circ\cdots\circ }^n)=:E_n,
\end{equation}
from which, one obtains
\begin{equation}\label{2b}
P(\bullet\overbrace{\circ\circ\cdots\circ }^n)=
P(\overbrace{\circ\circ\cdots\circ }^n\bullet)=E_n-E_{n+1},
\end{equation}
where $\bullet$ ($\circ$) indicates an occupied (empty) site, and
$P$ denotes the probability of the configuration. The evolution
equation for $E_n(t)$ is
\begin{align}\label{3}
{{{\rm d}E_n}\over{{\rm
d}t}}=&r_5P(\bullet\overbrace{\bullet\circ\cdots\circ
}^n)+r_3P(\circ\overbrace{\bullet\circ\cdots\circ }^n)\nonumber\\
&+r_6P(\overbrace{\circ\cdots\circ\bullet }^n\bullet)
+r_2P(\overbrace{\circ\cdots\circ\bullet }^n\circ)\nonumber\\
&-(r_1+r_2)P(\bullet\overbrace{\circ\circ\cdots\circ }^n)
-(r_3+r_4)P(\overbrace{\circ\cdots\circ\circ }^n\bullet).
\end{align}
The right-hand side of \Ref{3} is not generally in terms of only
$E_n$'s. In order to make it expressible in terms of $E_n$'s,
using \Ref{2b}, one has to impose
\begin{equation}\label{3b}
r_3=r_5,\qquad r_2=r_6.
\end{equation}
Then, using
\begin{equation}
P(\bullet\overbrace{\bullet\circ\cdots\circ
}^n)+P(\circ\overbrace{\bullet\circ\cdots\circ
}^n)=P(\overbrace{\bullet\circ\cdots\circ }^n)=E_{n-1}-E_{n},
\end{equation}
and another similar relation, one arrives at
\begin{equation}\label{4}
  {{\rm d}E_n(t)\over {\rm
  d}t}=(r_2+r_3)(E_{n-1}+E_{n+1}-2E_{n})-
  (r_1+r_4)(E_{n}-E_{n+1}),\qquad n>1.
\end{equation}
Note that if there were reactions with the final configuration
$\circ\circ$ then one encounters with terms like
$P(\circ\circ\bullet\circ\cdots \circ)$ or
$P(\circ\circ\bullet\bullet\circ\cdots \circ)$ at the right-hand
side of \Ref{3}, which are not expressible in terms of $E_{k}$'s.
On the other hand, if $\circ\circ$ is the initial configuration in
a reaction, the evolution equation for $E_n$'s is still closed,
although the analogue of \Ref{4} will be a linear finite
difference equation with nonconstant coefficients.

The equation of motion of $E_1(t)$ is
\begin{equation}\label{4b}
  {{\rm d}E_1(t)\over {\rm
  d}t}=(r_2+r_3)(1+E_2-2E_1)-
  (r_1+r_4)(E_1-E_2).
\end{equation}
It is seen that it takes a form similar to \Ref{4}, provided one
defines
\begin{equation}\label{4c}
E_0(t):=1.
\end{equation}
Then we have \Ref{4}, for $n\geq 1$, equipped with the boundary
condition \Ref{4c}. We also set $E_{L+1}(t)=0$, which means that
initially at least one particle is present in the lattice. If
initially all the sites were empty ($E_n(0)=1$ for all $n$), then
the above-defined reactions would not change the configuration of
the system, and if initially at least one particle was present,
then at any time $E_{L+1}(t)=0$. So, the completely empty lattice
is a stationary state which is decoupled from any other state.
Defining
\begin{equation}\label{3a}
b:={r_1+r_4\over r_2+r_3},
\end{equation}
and rescaling the time properly, the equation of motion becomes
\begin{equation}\label{4a}
  {{\rm d}E_n(t)\over {\rm
  d}t}=E_{n-1}+E_{n+1}-2E_{n}+
  b(E_{n+1}-E_{n}),\quad 0<n<L+1
\end{equation}
with the boundary conditions
\begin{equation}
E_0(t)=1,\qquad E_{L+1}(t)=0.
\end{equation}

A particular solution to this is the stationary solution:
\begin{equation}\label{4aa}
E^{\rm P}_{n-1}+E^{\rm P}_{n+1}-2E^{\rm P}_{n}+ b(E^{\rm
P}_{n+1}-E^{\rm P}_{n})=0.
\end{equation}
Taking the ansatz
\begin{equation}\label{5}
E^{\rm P}_n=A z_1^n +Bz_2^n,
\end{equation}
for $E_n^{\rm P}$, one arrives at
\begin{equation}\label{6}
z_i+z_i^{-1}-2+b(z_i-1)=0,
\end{equation}
the solutions of which are $z_1=1/(1+b) $ and $z_2=1$. Using the
boundary conditions $E_0=1$ and $E_{L+1}=0$, $A$ and $B$ are
obtained as
\begin{align}\label{7}
A&={1\over 1-(1+b)^{-L-1}},\nonumber \\
&\nonumber\\
B&={-(1+b)^{-L-1}\over 1-(1+b)^{-L-1}}.
\end{align}
Defining
\begin{equation}\label{8}
F_n(t):= E_n(t)-E_n^{\rm P},
\end{equation}
it is seen that the evolution equation for $F_n$ is the same as
that of $E_n$, but the boundary conditions for $F_n$ are
homogeneous. The initial condition for $F_n$ is
\begin{equation}\label{9}
F_n(0)=E_n(0)-E_n^{\rm P}.
\end{equation}
To calculate $F_n(t)$, one seeks the eigenvalues and eigenvectors
of the operator at the right-hand side of \Ref{4a}, that is
\begin{equation}\label{11}
\epsilon\;f_n=f_{n+1}+f_{n-1}-2f_n+b[f_{n+1}-f_n].
\end{equation}
The solution to this is
\begin{equation}\label{12}
f_n=a\; z_1^n+b\; z_2^n,
\end{equation}
where $z_i$'s satisfy
\begin{equation}\label{14}
z_i^2(1+b)-z_i(\epsilon+2+b)+1=0.
\end{equation}
Now, defining
\begin{equation}\label{15}
Z_i:=z_i\sqrt{1+b},
\end{equation}
it is seen $Z_1Z_2=1$. So
\begin{equation}\label{16}
f_n={1\over (1+b)^{n\over 2}}(a\; Z^n+b\; Z^{-n}).
\end{equation}
The boundary conditions $F_0(t)=F_{L+1}(t)=0$, lead to
\begin{equation}
Z=\exp\left({{i\pi k}\over{L+1}}\right),
\end{equation}
where $k$ is an integer satisfying $1<k<L+1$, and
\begin{equation}\label{18}
f_{k,n}={1\over{(1+b)^{n/2}}}\sin\left({{\pi
nk}\over{L+1}}\right).
\end{equation}
The corresponding eigenvalue is then
\begin{equation}\label{19}
\epsilon_k=-2-b+2\sqrt{1+b}\cos\left({{\pi nk}\over{L+1}}\right).
\end{equation}
Then $F_n(t )$ will be
\begin{equation}\label{21}
F_n(t)=\sum_{k=1}^L{{\alpha_k}\over{(1+b)^{n/2}}}\sin\left({{\pi
nk}\over{L+1}}\right)e^{E_k t},
\end{equation}
where
\begin{equation}\label{22}
\alpha_k={2\over{L+1}}\sum_{m=1}^L[E_m(0)-E_m^{\rm
P}](1+b)^{m/2}\sin \left({{m\pi k}\over{L+1}}\right).
\end{equation}

In the thermodynamic limit ($L\to \infty$), $F_n(t)$ takes a
simpler form. Defining $x:=\pi k/(L+1)$, \Ref{21} and \Ref{22}
lead to
\begin{align}\label{24}
  F_n(t)=&{2\over \pi} \sum_{m=1}^\infty \int_0^\pi {\rm d}x\;
  e^{(-2-b+2\sqrt{1+b}\cos x)t} \sin (nx) \sin (mx)F_m(0)
  (1+b)^{(m-n)/2}\nonumber\\
  &\nonumber\\
  =&\sum_{m=1}^\infty  (1+b)^{(m-n)/2} e^{-(2+b)t}
  ({\rm I}_{m-n}(2t\sqrt{1+b})-{\rm I}_{m+n}(2t\sqrt{1+b}))\nonumber\\
  &\times[E_n(0)-E_n^{\rm P}],
\end{align}
where in the second line we have used the integral representation
of the modified Bessel functions
\begin{equation}\label{36}
{\rm I}_n(t)={1\over \pi} \int_0^\pi {\rm d}x\cos (nx) e^{t\cos
x}.
\end{equation}

To study the large-time behaviour of $F_n(t)$, one takes $A_k=:
\pi B_k/(L+1)$. In the thermodynamic limit ($L\to \infty $),
\Ref{21} leads to
\begin{equation}\label{e36}
  F_n(t)={1\over (1+b)^{n/2}}\int_0^\pi {\rm d}x
  B(x)e^{(-2-b+2\sqrt{1+b}\cos x)t}\sin(nx) .
\end{equation}
At large times, the main contribution to the integral comes from
the region $x\approx 0$, in which the exponent of the exponential
term takes its largest value. So,
\begin{equation}\label{ee36}
  F_n(t)\approx{n\over (1+b)^{n/2}}e^{(-2-b+2\sqrt{1+b})t}
  \int_0^\pi {\rm d}x\;
 e^{-\sqrt{1+b} x^2t}x\; B(x),
\end{equation}
or
\begin{equation}\label{37}
F_n(t)\sim {e^{(-2-b+2\sqrt{1+b})t}\over t},
\end{equation}
provided $B(x)$ is well-behaved and nonzero at $x=0$. If $b\ne 0$,
there is an energy gap in the spectrum and the system relaxes
towards its stationary state exponentially. If $b=0$, there is no
energy gap and the relaxation towards the stationary state is in
the form of power law with the exponent $-1$.

The empty-interval probability functions can be used to obtain
some kinds of $n$-point functions. It is easy to see that
\begin{equation}
P(\bullet\overbrace{\circ\circ\cdots\circ }^m\bullet)=E_m-2
E_{m+1}+E_{m+2}.
\end{equation}
So, the results for $E_n(t)$'s can be used to obtain the
probability that between two occupied sites, there are $n$ sites,
which are empty.

\section{Models solvable through the empty interval method:
the general case} In the previous section, we considered
translationally-invariant initial conditions. As the dynamics is
translationally-invariant, the probability $E_n(t)$ will be the
same for all sites, provided the initial condition for it is so.
In this section, we release the translational invariance of the
initial conditions and the quantity of our interest is the
probability $E_{k,n}(t)$, that $n$ consecutive sites, starting
from the site $k$ are empty at time $t$:
\begin{equation}
P_k(\overbrace{\circ\circ\cdots\circ }^n)=E_{k,n}
\end{equation}
It is easy to see that
\begin{align}\label{38}
P_l(\bullet\overbrace{\circ\circ\cdots\circ }^m)&=E_{l+1,m}-E_{l,m+1},
\nonumber \\
P_l(\overbrace{\circ\circ\cdots\circ }^m
\bullet)&=E_{l,m}-E_{l,m+1}.
\end{align}
Using the interactions \Ref{e1}, and the above identities, one
arrives at
\begin{align}\label{39}
{{\rm d}E_{k,n}(t)\over {\rm
 d}t}=&r_3(E_{k+1,n-1}-E_{k,n})+r_2(E_{k,n-1}-
 E_{k,n})\nonumber \\ & -(r_1+r_2)(E_{k,n}-
 E_{k-1,n+1})-(r_3+r_4)(E_{k,n}-E_{k,n+1}).
\end{align}
Similar to the previous section, imposing the boundary condition
\begin{equation}\label{40}
E_{k,0}(t)=1,
\end{equation}
makes the evolution equation valid for $0<n<L+1$. If the lattice
is not initially empty, one also has
\begin{equation}\label{40b}
E_{k,L+1}(t)=0.
\end{equation}
Using the definitions
\begin{align}\label{41}
b:=&{r_1+r_4\over r_2+r_3},\nonumber \\
c:=&{r_1+r_2\over r_2+r_3},\nonumber \\
d:=&{r_3\over r_2+r_3},
\end{align}
equation \Ref{39} can be rearranged in the form
\begin{align}\label{42}
{{\rm d}E_{k,n}(t)\over {\rm d}t}=& E_{k,n-1}+E_{k,n+1}-2E_{k,n}-
b(E_{k,n}-E_{k,n+1})
\nonumber\\
       &- c(E_{k,n+1}-
       E_{k-1,n+1})+d(E_{k+1,n-1}-E_{k,n-1}).
\end{align}
Using the particular solution $E^{\rm P}_n$, one defines
\begin{equation}\label{43}
F_{k,n}(t):=E_{k,n}(t)-E^{\rm P}_n,
\end{equation}
which satisfies \Ref{42}, but with homogeneous boundary
conditions:
\begin{equation}\label{44}
F_{k,0}(t)=F_{k,L+1}(t)=0.
\end{equation}
Applying the Fourier transformation
\begin{equation}\label{45}
  \tilde F_n(\omega ,t):=\sum_k \omega^k F_{k,n}(t),
\end{equation}
one arrives at
\begin{align}\label{45c}
{{\rm d}\tilde F_n(\omega,t)\over {\rm d}t}=& \tilde
F_{n-1}(\omega,t)+\tilde F_{n+1}(\omega,t)-2\tilde F_n(\omega,t)-
b[\tilde F_n(\omega,t)-\tilde F_{n+1}(\omega,t)]
\nonumber\\
       &- c[\tilde F_{n+1}(\omega,t)-
       \omega\tilde F_{n+1}(\omega,t)]+d[\omega^{-1}\tilde F_{n-1}(\omega,t)
       -\tilde F_{n-1}(\omega,t)].
\end{align}
To solve this, one first solves the eigenvalue problem
\begin{align}\label{45d}
\epsilon\tilde f_n(\omega)=& \tilde f_{n-1}(\omega)+\tilde
f_{n+1}(\omega)-2\tilde f_n(\omega)- b[\tilde f_n(\omega)-\tilde
f_{n+1}(\omega)]
\nonumber\\
       &- c[\tilde f_{n+1}(\omega)-
       \omega\tilde f_{n+1}(\omega)]+d[\omega^{-1}\tilde f_{n-1}(\omega)
       -\tilde f_{n-1}(\omega)].
\end{align}
using the ansatz
\begin{equation}
\tilde f_n=a\; z_1^n+b\; z_2^n,
\end{equation}
and the boundary conditions, one arrives at
\begin{align}\label{47}
\tilde f_{s,n}(\omega)&=[B(\omega)]^{n/2}\sin\left({{n\pi
s}\over{L+1}}\right)\nonumber\\
\epsilon_s&= -2-b+D(\omega )\cos\left({{\pi s}\over{L+1}}\right),
\end{align}
where
\begin{align}
D(\omega ):=& 2\sqrt{[1+d(\omega^{-1}-1)][1+b+c(\omega
-1)]},\nonumber \\
B(\omega):=&{1+d(\omega^{-1}-1)\over 1+b+c(\omega -1)},
\end{align}
and $s$ is an integer between $1$ and $L$. In the thermodynamic
limit $L\to\infty$, one arrives at
\begin{align}\label{50}
\tilde F_n(\omega ,t)=&\sum_{m}\tilde F_m(\omega
,0)[B(\omega)]^{(n-m)/2} e^{-(2+b)t}\nonumber\\
&\times\{{\rm I}_{m-n}[D(\omega)t]-{\rm I}_{m+n}[D(\omega)t]\},
\end{align}
where
\begin{equation}\label{51}
\tilde F_m(\omega ,0)=\sum_k\omega^k [E_{k,m}(0)-E_{m}^{\rm P}].
\end{equation}

Now, let's consider the relaxation of the system towards its
stationary state. Suppose that the initial value for $E_{k,n}$ is
so that $\tilde F_n(\omega,0)$ contains a term proportional to
$\delta(p)$ (where $\omega=e^{ip}$), and another term which is a
smooth function of $\omega$. The delta term comes from a
translationally-invariant part in $F_{n,k}(0)$. Using the steepest
descent method, one can see that the relaxation behavior of the
second term is governed by the extremum value of the eigenvalues
$\epsilon$ with respect to a complex $\omega$. This is found to be
\begin{equation}\label{53}
\epsilon_{{\rm max}}=-2-b+2[\sqrt{dc}+\sqrt{(1-d)(1+b-c)}].
\end{equation}
It is easy to show
\begin{equation}\label{54}
\epsilon_{{\rm max}}\leq -2-b+2\sqrt{1+b}.
\end{equation}
Equality holds when
\begin{equation}
r_2(r_1+r_2)=r_3(r_3+r_4).
\end{equation}
This means that the relaxation time for the
translationally-noninvariant part is smaller than of the
translationally-invariant part. That is, the
translationally-noninvariant fluctuations disappear faster than
the translationally invariant parts.
\section{The continuum limit}
In the previous section, we considered the probability of finding
$n$ consecutive empty sites starting from the $k$-th site. For the
continuum limit, it is better to use a quantity with arguments
symmetric relative to the starting point and the end point of the
empty interval. For an empty interval of the length $n$ starting
from the site $k$, the end site is $k'=2k+n-1$. Then one can use
\begin{equation}\label{55}
s:=k+k'=2k+n-1
\end{equation}
instead of $k$ for labeling the empty interval. $s/2$ is the
center of the empty interval. So, one uses the quantity
\begin{equation}
{\cal E}_{s,n}(t):=E_{k,n}(t).
\end{equation}
The equation of motion for ${\cal F}_{s,n}(t)$ is then
\begin{align}\label{56}
{{\rm d}{\cal F}_{s,n}(t)\over {\rm
 d}t}=&r_2({\cal F}_{s-1,n-1}+{\cal F}_{s-1,n+1}-
      2{\cal F}_{s,n})+r_3({\cal F}_{s+1,n-1}+
      {\cal F}_{s+1,n+1}-2{\cal F}_{s,n})\nonumber \\
       & +r_4({\cal F}_{s+1,n+1}-{\cal F}_{s,n})
       +r_1({\cal F}_{s-1,n+1}-{\cal F}_{s,n}).
\end{align}
Here ${\cal F}$ is the solution to the evolution equation of
${\cal E}$, but with homogeneous boundary conditions. Using
$X:=s/2$ and $x:=n$, and Taylor-expanding the above expression in
the continuum limit, one arrives at
\begin{equation}\label{57}
 {{\partial{\cal F}(X,x;t)}\over{\partial t}}=(A\partial_X+B\partial_x+
            {C\over 4}\partial^2_X+C\partial^2_x+
            D\partial_x\partial_X){\cal F}(X,x;t),
\end{equation}
where the parameters $A$, $B$, $C$, and $D$ are
\begin{align}\label{58}
A:=r_3-r_2+{r_4-r_1\over 2},& \qquad B:=r_1+r_4\nonumber\\
C:={1\over 2}[r_1+r_4+2(r_2+r_3)],&\qquad D:={r_4-r_1\over 2}.
\end{align}
Using the change of variables
\begin{equation}\label{59}
\hat X:=X+(A-{BD\over 2C})t-{D\over 2C}x,
\end{equation}
and
\begin{equation}\label{60}
{\cal F}(X,x;t)=: \exp [-{B\over 2C}x-{B^2\over 4C}t]
 \hat {\cal F}(\hat X,x;t),
\end{equation}
one arrives at
\begin{equation}\label{61}
{{\partial\hat{\cal F}(\hat X,x;t)}\over{\partial
t}}=[C\partial^2_x +({C\over 4}-{D^2\over 4C})\partial^2_{\hat
X}]\hat{\cal F}(\hat X,x;t).
\end{equation}
The boundary conditions for  $\hat{\cal F}$ are
\begin{equation}\label{62}
{\hat{\cal F}}(\hat X,x=0;t)=\hat{\cal F}(\hat X,x\to \infty
;t)=0.
\end{equation}
The Green function $G(\hat X, \hat X',x,x';t)$ for the equation
\Ref{61} with the boundary conditions \Ref{62} is
\begin{align}\label{63}
G(\hat X, \hat X',x,x';t):=&{1\over 4\pi t\sqrt{CC'}}e^{-(\hat X-
\hat X')/(4C't)}\nonumber\\
&\times\left[e^{-(x- x')/(4Ct)}-e^{-(x+x')/(4Ct)}\right],
\end{align}
where $C':=(C/4)-D^2/(4C)$. Finally,
\begin{align}\label{64}
{\cal F}(X,x;t)=&\exp\left(-{B\over 2C}x-{B^2\over 4C}t\right)
\nonumber\\
&\times \int_{x'=0}^\infty\int_{X'=-\infty}^\infty {\rm d}x'{\rm
d}X'\;G(\hat X, \hat X',x,x';t){{\cal F}}(X',x';t)e^{(B/2C)x'}.
\end{align}
To obtain the solution for ${\cal E}$, one has to add a particular
solution with the boundary conditions
\begin{equation}
{\cal E}(x=0)=1,\qquad {\cal E}(x\to\infty)=0.
\end{equation}
One can choose this particular solution to be
translationally-invariant (that is $X$-independent). One is then
led to
\begin{equation}\label{65}
(B\partial_x+C\partial^2_x ){\cal E}^{\rm P}(x)=0,
\end{equation}
the solution to which is
\begin{equation}
{\cal E}^{\rm P}=\exp(-Bx/C)
\end{equation}
This particular solution is the same as that of section 2 in the
continuum limit.

\vskip\baselineskip

\noindent {\bf Acknowledgement} \\The authors would like to thank
Institute for Studies in Theoretical Physics and Mathematics for
partial support. M. Alimohammadi would like to thank also the
research council of the University of Tehran, for partial
financial support.

\end{document}